\documentstyle[fleqn,espcrc3]{article}

% ...
\newcommand{\be}{\begin{equation}}
\newcommand{\ee}{\end{equation}}
\newcommand{\bea}{\begin{eqnarray}}
\newcommand{\eea}{\end{eqnarray}}

\newcommand{\dalpha}{\dot{\alpha}}
\newcommand{\dbeta}{\dot{\beta}}

\newcommand{\bD}{\bar{D}}
\newcommand{\cD}{{\cal D}}
\newcommand{\cbD}{\bar{{\cal D}}}

\newcommand{\bQ}{\bar{Q}}
\newcommand{\blambda}{\bar{\lambda}}
\newcommand{\bJ}{\bar{J}}

\newcommand{\nn}{\nonumber}

\newcommand{\AmS}{{\protect\the\textfont2
 A\kern-.1667em\lower.5ex\hbox{M}\kern-.125emS}}

\title{Partial Breaking of Extended Supersymmetry}

\author{Jonathan A.  Bagger
\address{Department of Physics and Astronomy,
 Johns Hopkins University \\
 3400 N.  Charles Street,
 Baltimore, MD 21218}
 \thanks{Supported by the U.S.  National
Science Foundation, grant NSF-PHY-9404057.}
 }

\begin{document}

\begin{abstract}
\noindent
In this talk we use nonlinear realizations to study the
spontaneous partial breaking of rigid and local\newline
supersymmetry.
\end{abstract}

\maketitle

\section{INTRODUCTION}

In this talk we will consider the partial breaking of extended
supersymmetry.  For simplicity, we will restrict our attention
to the case $N=2 \rightarrow N=1$, but much of what we find
can be readily extended to the case of higher supersymmetries,
spontaneously broken to $N=1$.

There is a heuristic argument which implies that extended
supersymmetry cannot be spontaneously broken to $N=1$ in
four dimensions \cite{Witten}.  The argument runs as follows:
Suppose that there are two supersymmetries, one broken and
one unbroken.  Since one supersymmetry is preserved, one
supercharge must annihilate the vacuum.  If the Hilbert space
is positive definite, the supersymmetry algebra
\be
\{\,Q_\alpha^A, \bQ_{\dalpha B}\} = 2 \sigma^m_{\alpha\dalpha}
\,P_m\,\delta^A{}_B
\label{susy alg}
\ee
implies that the Hamiltonian must also annihilate the vacuum.
This, in turn, requires the second supercharge to annihilate
the vacuum, so the other supersymmetry cannot be broken.

Hughes, Liu and Polchinski \cite{HLP} found a legal loophole
which allowed them to evade this argument.  They exploited the
fact that in theories with spontaneous symmetry breaking, the
broken symmetry charges do not exist.  This motivated them to
consider the following current algebra,
\bea
\{\,Q_\alpha^1, \bJ_{m \dalpha 1}\} &=& 2 \sigma^n_{\alpha\dalpha}
\,T_{mn} \nn\\
\{\,Q_\alpha^2, \bJ_{m \dalpha 2}\} &=& 2 \sigma^n_{\alpha\dalpha}
\,(v^4 \eta_{mn} + T_{mn}).
\label{current alg}
\eea
Note that the right-hand sides of the two commutators differ
by a constant.  In the limit $v \rightarrow 0$,
the constant vanishes and the
current algebra can be integrated to give the charge algebra
(\ref{susy alg}).
When $v \ne 0$, the current algebra cannot be integrated,
and the second supersymmetry is spontaneously broken.  (Of
course, if there were just one supersymmetry, the constant
would become the vacuum energy that signals spontaneous
supersymmetry breaking.)

Hughes, Liu and Polchinski found an explicit realization of
their algebra in terms of a four-dimensional supermembrane
propagating in six-dimensional superspace.  They found its
invariant action and demonstrated that it realizes
the partial breaking of extended supersymmetry.

The membrane approach leaves many open questions, some
of which will be addressed in this talk.  In particular, we
would like to know if there are other realizations of partial
supersymmetry breaking. We would like to know whether the
$N=2$ supersymmetry gives rise to any restrictions on the $N=1$
matter couplings.  And we would like to know if the system can
be coupled to supergravity, because gravity can
distinguish between the different stress-energy tensors on
the right-hand side of eq.~(\ref{current alg}).  (For
alternative approaches to this subject, and other references,
see \cite{APT et al} -- \cite{others 2}, and references therein.)

\section{COSET CONSTRUCTION}

In this talk we will take a bottom-up approach to the subject
of partial supersymmetry breaking.  We will use nonlinear
realizations to describe the effective $N=1$ theory which holds
below the scale of the second supersymmetry breaking.  We will
use the formalism of Coleman, Wess and Zumino \cite{CWZ}, as
modified by Volkov \cite{Volkov}, to construct theories where
the $N=1$ supersymmetry is manifest, and the second supersymmetry
is nonlinearly realized.

The approach of Coleman, Wess, Zumino and Volkov is based on a
coset decomposition of a symmetry group, $G$.  We start with
a group, $G$, of internal and spacetime symmetries, and
partition the generators of $G$ into three classes:
\begin{itemize}
\item
$\Gamma_A$, the generators of unbroken spacetime translations;
\item
$\Gamma_a$, the generators of spontaneously broken internal and
spacetime symmetries; and
\item
$\Gamma_i$, the generators of unbroken spacetime rotations
and unbroken internal symmetries.
\end{itemize}
The generators $\Gamma_i$ close into the stability group, $H$.

Given $G$ and $H$, we define the coset $G/H$ in terms of
an equivalence relation on the elements $\Omega \in G$,
$\Omega \sim \Omega\, h$, with $h \in H$.  The coset can be
thought of as a section of a fiber bundle with total space,
$G$, and fiber, $H$.

This equivalence relation suggests that we parametrize
the coset as follows,
\begin{equation}
\Omega = \exp i X^A \Gamma_A\ \exp i\xi^a(X) \Gamma_a.
\end{equation}
Physically, the $X^A$ play the role of generalized spacetime
coordinates, while the $\xi^a(X)$ are generalized Goldstone
fields, defined on the generalized coordinates and valued in
the set of broken generators $\Gamma_a$.  There is one
generalized coordinate for every unbroken spacetime translation,
and one generalized Goldstone field for every spontaneously
broken generator.

We define the action of the group $G$ on the coset $G/H$ by
left multiplication, $\Omega \rightarrow g\, \Omega = \Omega^\prime
\, h$, with $g \in G$.  In this expression,
\begin{equation}
\label{define primes}
\Omega^\prime = \exp i X^{\prime A} \Gamma_A
\ \exp i \xi^{\prime a}(X^\prime) \Gamma_a
\end{equation}
and $h = \exp i \alpha^i (g, X, \xi) \Gamma_i.$
The group multiplication induces nonlinear transformations
on the coordinates $X^A$ and the Goldstone fields $\xi^a$:
\be
X^A \rightarrow X^{\prime A}, \quad
\xi^a(X) \rightarrow \xi^{\prime a}(X^\prime).
\ee
These transformations realize the full symmetry group, $G$.  Note
that the field $\xi^a$ transforms by a shift under the transformation
generated by $\Gamma_a$.  This confirms that $\xi^a$ is indeed the
Goldstone field corresponding to the broken generator $\Gamma_a$.

An arbitrary $G$ transformation induces a compensating $H$
transformation which is required to restore the section.  This
transformation can be used to lift any representation, $R$, of
$H$, to a nonlinear realization of the full group, $G$, as
follows,
\begin{equation}
\chi(X) \rightarrow \chi^\prime(X^\prime) = D(h)
\chi(X).
\end{equation}
Here $D(h) = \exp(i \alpha^i T_i)$, where $\alpha^i$ was defined
below (\ref{define primes}), and the $T_i$ are generators of $H$ in
the representation $R$.

To construct an invariant action, it is helpful to have a
vielbein, connection and covariant derivative, built from
the Goldstone fields in the following way.  One first computes
the Maurer-Cartan
form, $\Omega^{-1} d\Omega$, where $d$ is the exterior derivative.
One then expands $\Omega^{-1} d\Omega$ in terms of the Lie algebra
of $G$,
\begin{equation}
\Omega^{-1} d \Omega = i (\omega^A \Gamma_A +
\omega^a \Gamma_a + \omega^i\Gamma_i),
\label{omegas}
\end{equation}
where $\omega^A, \omega^a$ and $\omega^i$ are one-forms on the
manifold parametrized by the coordinates $X^A$.

The Maurer-Cartan form transforms as follows under a rigid
$G$ transformation,
\begin{equation}
\Omega^{-1} d \Omega \rightarrow h (\Omega^{-1} d \Omega) h^{-1}
 - dh\,h^{-1}.
\end{equation}
{}From this we see that the fields $\omega^A$ and $\omega^a$
transform covariantly under $G$, while $\omega^i$ transforms by
a shift.  These transformations help us identify
\begin{equation}
\omega^A = dX^M \, E_M{}^A
\end{equation}
as the covariant vielbein,
\begin{equation}
\omega^a = dX^M \, E_M{}^A \cD_A \xi^a
\end{equation}
as the covariant derivative of the Goldstone field $\xi^a$,
and
\begin{equation}
\omega^i = dX^M \, \omega_M^i
\label{h connection}
\end{equation}
as the connection associated with the stability group, $H$.
With these building blocks, it is easy to construct an
action invariant under the full group $G$.

The coset construction is very general and very powerful.
For the case of internal symmetries, it allows one to prove
that any $H$-invariant action can be lifted to be $G$-invariant
with the help of the Goldstone bosons.  For $N=1$ supersymmetry,
it can be used to show that any Lorentz-invariant action can
be made supersymmetric with the help of the Goldstone
fermion.

\section{NONLINEAR SUPERSYMMETRY}

In this section we will show that any $N=1$ supersymmetric
theory can be made $N=2$ supersymmetric with the help of an
$N=1$ Goldstone superfield.  We will find that the Goldstone
superfield can contain either an $N=1$ chiral or vector multiplet.
Note, however, that the coset construction does not tell us
anything about the underlying theory in which both
supersymmetries are linearly realized.  Indeed, such a theory
is not even guaranteed to exist.

To begin, let us rewrite the $N=2$ supersymmetry algebra as
follows,
\bea
\{Q_\alpha, \bar Q_{\dot\alpha}\} =
2\sigma^a_{\alpha\dot\alpha}P_a, \;\;
\{S_\alpha, \bar S_{\dot\alpha}\}=
2\sigma^a_{\alpha\dot\alpha}P_a, &&\nn\\
\{Q_\alpha, S_\beta\} = 2\epsilon_{\alpha\beta}Z, \;\;
\{\bar Q_{\dot\alpha}, \bar S_{\dot\beta}\}=
2\epsilon_{\dot\alpha\dot\beta}\bar Z.\;\;\;\;\,&&
\label{susy alg 2}
\eea
Here $Q_\alpha$ and $S_\alpha$ are the supersymmetry generators,
$P_a$ the four-dimensional momentum operator, and $Z$ is a complex
central charge.  In what follows, we will define $Q_\alpha$ to be
the unbroken $N=1$ supersymmetry generator, and $S_\alpha$ to be
its broken counterpart.

We shall first take a minimal approach, and choose the group $G$
to be the supergroup whose algebra is (\ref{susy alg 2}).
We will take the subgroup $H$ to be the supergroup generated
by $P_a$, $Q_\alpha$ and $\bQ_{\dalpha}$.  We parametrize the coset
element $\Omega$ as follows,
\bea
\label{real_parametrization}
\Omega &=& \exp i(x^aP_a +\theta^\alpha Q_\alpha +\bar\theta_{\dot\alpha}
\bar Q^{\dot\alpha})\nn\\
&&\quad\times
\exp i(\psi^\alpha S_\alpha +\bar\psi_{\dot\alpha}\bar S^{\dot\alpha}).
\eea
Here $x,\ \theta$ and $\bar\theta$ are the coordinates of $N=1$
superspace, while $\psi^\alpha$ and its conjugate $\bar\psi_{\dot\alpha}$
are Goldstone $N=1$ superfields of (geometrical) dimension $-1/2$.
These spinor superfields contain far too many component fields, so
we need to find a set of consistent, covariant constraints to reduce
the number of fields.

The correct constraints are most easily expressed in term of the
$N=2$ covariant derivatives of the Goldstone superfield.
These covariant derivatives can be explicitly written as follows,
\begin{eqnarray}
\label{derivatives}
{\cal D}_\alpha &=& D_\alpha
 - i(D_\alpha\psi\sigma^a\bar\psi +
 D_\alpha\bar\psi\bar\sigma^a\psi)\omega^{-1}_a{}^m\partial_m \nn \\
\bar{\cal D}_{\dot\alpha} &=& \bar D_{\dot\alpha} -
i(\bar D_{\dot\alpha}\psi\sigma^a\bar\psi +
\bar D_{\dot\alpha}\bar\psi\bar\sigma^a\psi)\omega^{-1}_a{}^m
\partial_m\nn\\
{\cal D}_a &=& \omega^{-1}_a{}^m\partial_m,
\end{eqnarray}
where $\omega_m{}^a \equiv \delta_m^a + i(\partial_m\psi\sigma^a\bar\psi
+\partial_m\bar\psi\bar\sigma^a\psi)$ and $D_\alpha,\ \bar D_{\dot\alpha}$
are ordinary flat $N=1$ superspace spinor derivatives.  The
covariant derivatives obey the following commutation relations,
\begin{eqnarray}
\label{cov_algebra}
\{ {\cal D}_\alpha, {\cal D}_\beta \} & = &
- 2i({\cal D}_\alpha\psi^\gamma
{\cal D}_\beta\bar\psi^{\dot\gamma} +
(\alpha \leftrightarrow \beta))
{\cal D}_{\gamma\dot\gamma} \nn\\
\left[ {\cal D}_\alpha, {\cal D}_a \right] & = &
-2i ({\cal D}_\alpha \psi^\gamma
{\cal D}_a \bar\psi^{\dot\gamma} +
(\alpha\leftrightarrow a))
{\cal D}_{\gamma\dot\gamma} \nn \\
\{ {\cal D}_\alpha, \bar{\cal D}_{\dot\beta} \} & = &
2i\sigma^a_{\alpha\dot\beta}{\cal D}_a
-2i({\cal D}_\alpha\psi^\gamma
\bar{\cal D}_{\dot\beta}\bar\psi^{\dot\gamma} \nn\\
&& \quad + (\alpha \leftrightarrow \dot\beta ))
{\cal D}_{\gamma\dot\gamma} ,
\end{eqnarray}
where ${\cal D}_{\alpha \dalpha} \equiv \sigma^a_{\alpha \dalpha}
{\cal D}_a$.

The first set of constraints is simply \cite{BG1}
\bea
\cbD\cbD\,\psi_\alpha &=& {\cal O}(\psi^3) \nn\\
\cD_\alpha \psi_\beta + \cD_\beta \psi_\alpha
 &=& {\cal O}(\psi^3).
\label{chiral constraints}
\eea
The right-hand side of this equation must be adjusted for
consistency with (\ref{cov_algebra}).  Remarkably, this can
be done using the dimensionless invariants
$\bar{\cal D}_{\dot\alpha} \psi_\alpha$ and
${\cal D}_\alpha\psi_\beta$
(together with their complex conjugates).  It turns
out that there is a unique, consistent solution order-by-order
in powers of the Goldstone field.

The solution to the constraints (\ref{chiral constraints}) is
easy to find in perturbation theory.  To lowest order, it is just
the chiral multiplet $\phi$,
\bea
\psi_\alpha &=& D_\alpha \phi + {\cal O}(\psi^3)  \nn\\
\bD_{\dalpha} \phi &=& {\cal O}(\psi^3).
\eea
In this expression, $D_\alpha$ is the ordinary $N=1$ superspace
spinor derivative.

The second set of constraints is \cite{BG2}
\bea
\cbD_{\dalpha} \psi_\alpha &=& {\cal O}(\psi^3) \nn\\
\cD^\alpha \psi_\alpha + \cbD_{\dbeta} \bar\psi^{\dbeta}
 &=& {\cal O}(\psi^3).
\label{vector constraints}
\eea
As above, the right-hand side must be adjusted for
consistency with the algebra of covariant derivatives.
Again, there is a unique, consistent solution.  To
lowest order in perturbation theory, it is just
\bea
\psi_\alpha &=& W_\alpha + {\cal O}(\psi^3)  \nn\\
W_\alpha &=& - {1\over4}\bD\bD D_\alpha V + {\cal O}(\psi^3),
\eea
where $V$ is a real $N=1$ vector superfield.  We see
that the chiral and vector Goldstone multiplet can each
be obtained to lowest order in perturbation theory.
In fact, the consistency of the multiplets survives
to all orders in perturbation theory.

The Goldstone action can be constructed order-by-order in the
Goldstone fields.  For the chiral case, it is simply \cite{BG1}
\be
\label{goldaction1}
S = v^4 \int d^4x d^2\theta d^2\bar\theta \,E\,
[\phi^+ \phi + {\cal O}(\phi^4)].
\ee
In this expression, $E =$Ber($E_{M}{}^{A}$) is the superdeterminant
of the vielbein, and $v$ is the constant of dimension one which
corresponds to the scale of the supersymmetry breaking.
The action (\ref{goldaction1}) is invariant under the full $N=2$
supersymmetry.

For the vector multiplet, the Goldstone action is just \cite{BG2}
\bea
\label{goldaction2}
S &=& {v^4\over 4} \int d^4x d^2 \theta \,{\cal E}\, W^2 + h.c. \nn\\
&& \quad + \int d^4x d^4\theta \,E\, {\cal O}(W^4).
\eea
This action is invariant under $N=2$ supersymmetry.  It is also
gauge-invariant.  Curiously enough, the gauge field contribution to
the Goldstone action coincides with the expansion of the Born-Infeld
action.

Having constructed the $N=2$ Goldstone action, we are
now ready to add $N=2$ covariant matter.  The basic ingredients
are $N=2$ nonlinear generalizations of $N=1$ chiral and vector
superfields.  The generalized chiral superfields are defined by
the constraint $\cbD_{\dalpha}\chi = 0$, while vector superfields
are defined by the reality condition $V = V^+$.
These constraints are consistent for either type of Goldstone
multiplet.

The matter action is easy to write down for either Goldstone
multiplet.  The kinetic term is
\be
S = \int d^4x d^4\theta  \,E \, K(\chi^+,\chi)
\ee
while the superpotential term is
\be
S = \int d^4x d^2\theta \,{\cal E}\,P(\chi).
\ee
As before, $E$ and ${\cal E}$ are superdeterminants of the
supervielbein $E_M{}^A$.  They can be adjusted to preserve the
condition
\be
\int d^4x d^4\theta \,E\, F(\chi) = 0 .
\ee
This allows the matter action to be K\"ahler invariant, so the
matter couplings are described in terms of K\"ahler manifolds,
just as for $N=1$.

It is not hard to generalize these results to include vector
superfields.  Our general conclusion is that any $N=1$ invariant
theory can be lifted to be $N=2$ supersymmetric with the help of a
Goldstone superfield.  Furthermore, we find that the Goldstone
superfield can contain either an $N=1$ chiral or vector multiplet.

Now that we have two explicit realizations of partial supersymmetry
breaking, we can ask how they avoid the no-go argument
discussed above.  In each case, the nonlinear theory exploits the
loophole of Hughes, Liu and Polchinski.  The second supercurrent
goes like $J^m_\alpha \sim v^4 \sigma^m_{\alpha\dalpha}
\blambda^{\dalpha}$, so its commutator with the second supercharge
reproduces the algebra (\ref{current alg}).

\section{GEOMETRY}

The fact that the constraints need to be adjusted order-by-order
in $\psi_\alpha$ hints that a deeper structure underlies
partial supersymmetry breaking.  The $N=2$ supersymmetry does not
provide enough symmetry to uniquely fix the covariant derivatives
and the associated constraints.  This intuition is borne
out for the case of the chiral multiplet, where a much deeper set
of symmetries acts on the Goldstone multiplet \cite{BG1}.

To see this, let us consider a coset where the group $G$ contains
not only $N=2$ supersymmetry, but also its maximal
automorphism group, $SO(5,1) \times SU(2)$, where the $SU(2)$
acts on the two supersymetry generators, and $SO(5,1)$ is the
$D=6$ Lorentz group.  (Under $SO(5,1)$, the generators $P_a$ and
$Z$ form a $D=6$ vector, while the supercharges form a $D=6$
Majorana-Weyl spinor).  Let us take $H$ to be $SO(3,1)\times
SO(2) \times U(1)$, where $SO(3,1)\times SO(2) \subset SO(5,1)$,
$U(1)\subset SU(2)$, and $SO(3,1)$ is the $D=4$ Lorentz group.

Our parametrization of the coset $G/H$ involves the $N=1$
superspace coordinates, as well as different Goldstone
superfields for each of the broken symmetries,
\bea
\Omega &=& \exp i(x^aP_a+\theta^\alpha Q_\alpha
+\bar\theta_{\dot\alpha}\bar Q^{\dot\alpha}) \nn\\
&&\times
\exp i(\Phi Z +\bar\Phi\bar Z +
\Psi^\alpha S_\alpha+\bar\Psi_{\dot\alpha}\bar S^{\dot\alpha}) \nn\\
&&\times \exp i(\Lambda^aK_a+ \bar\Lambda^a\bar K_a +\Xi T
+\bar\Xi\bar T).
\eea
Here $\Lambda^a,\, \bar\Lambda^a$ are the Goldstone superfields
associated with the generators $K_a,\, \bar K_a$ of $SO(5,1)/
SO(3,1)\times SO(2)$.  Similarly, $\Xi,\, \bar\Xi$ are the
Goldstone superfields for the broken generators
$T,\, \bar T$ of $SU(2)/U(1)$.

As before, the $N=1$ Goldstone superfields contain far more components
than the minimal Goldstone multiplet.  This motivates us to impose
the following consistent set of constraints:
\bea
&&{\bar{\cal D}}_{\dot\alpha}\Phi=0,
\ \ {\cal D}_{\alpha}\Phi=0,
\ \ {\cal D}_{a}\Phi=0 \nn \\
&&{\cal D}_{\alpha}\Psi^\beta=0,
\ \ {\bar{\cal D}}_{\dot\alpha}\Psi^\beta=0.  \label{constr2}
\eea
These constraints allow us to express the
Goldstone superfields $\Psi^\alpha, \Lambda^a$ and $\bar\Xi$ in
terms of a single superfield $\Phi$.  To lowest order,
we find $\Psi^\alpha = -{i\over 2}D^\alpha\Phi$,
$\Lambda_a = -\partial_a\Phi$, and
$\bar\Xi = {1\over 4}D^2\Phi$.
The constraint ${\bar{\cal D}}_{\dot\alpha}\Phi=0$ reduces
$\Phi$ to an $N=1$ chiral superfield.

The remarkable fact about this construction is that it reveals
a geometrical role for each component of the chiral Goldstone
multiplet.  The scalar field, $A$, is the complex Goldstone boson
associated with the spontaneously broken central charge symmetry.
Its derivative, $\partial_m A$, is the Goldstone boson associated
with $SO(5,1)/SO(3,1)\times SO(2)$.  The $F$-component of $\Phi$
is the complex Goldstone boson associated with the $SU(2)/U(1)$.
Finally, the spinor is the Goldstone fermion that arises
from the partially broken supersymmetry.

The action (\ref{goldaction1}) turns out to be invariant under
$SO(5,1)$, but it explicitly breaks $SU(2)$ down to $U(1)$.
Furthermore, any $R$-invariant $N=1$ matter action can
be made $SO(5,1)$ invariant.
These facts hint that the Goldstone action might be related
to the six-dimensional membrane of Hughes, Liu and Polchinski.
Indeed, it is not hard to show that the chiral Goldstone action
is precisely the gauge-fixed membrane action.

The geometry that underlies the vector case is presently under
study.  The Born-Infeld form of the gauge action suggests that
it might be related to some sort of D-brane in a
higher dimension.  But no matter what, one would like to find
the Goldstone-type symmetries associated with the gauge field
strength and the auxiliary field of the Goldstone multiplet.

In fact, the $D$-component of the Goldstone multiplet
can be interpreted as the Goldstone boson associated with
the following $U(1)$ subgroup of the $SU(2)$ automorphism
symmetry:
$\delta\theta^\alpha = i\lambda\psi^\alpha$,
$\delta\psi^\alpha = i\lambda\theta^\alpha$.
Under such a transformation, the $D$-component is shifted by
the constant parameter $\lambda$.

If we were to extend $G$ in $G/H$ by this $U(1)$, we would
eliminate the dimensionless invariant ${\cal D}^\alpha\psi_\alpha$
in favor of the corresponding Goldstone superfield.  Even then,
there would still be a dimensionless invariant associated with
the gauge field strength, ${\cal D}_{(\alpha}\psi_{\beta)}$.
This suggests that there is an extension of $N=2$ supersymmetry
which associates a Goldstone-like symmetry with this field
strength.

Moreover, gauge fields themselves can be interpreted as
Goldstone fields associated with infinite-dimensional symmetry
groups.  This leads us to wonder whether the full symmetry of
the new multiplet is some infinite-dimensional extension of
$N=2$ supersymmetry.

\section{SUPERGRAVITY}

We have just seen that there are two possible Goldstone realizations
of partial supersymmetry breaking.  Both give rise to the current
algebra (\ref{current alg}).  Because of the curious shift in the
stress-energy tensor, one would like to couple the Goldstone multiplets
to supergravity.  Presumably this is possible \cite{others 1},
\cite{others 2}, in which case we would like to study the current
algebra.

It is helpful to start our analysis by counting degrees of freedom.
Since we wish to couple to $N=2$ supergravity, we need to include the
$N=2$ supergravity multiplet, which contains one graviton, two
gravitinos, and one vector.  Moreover, since we are partially breaking
supersymmetry, we also need to include enough fields to make up
a massive $N=1$ spin-3/2 multiplet, which has one massive
spin-3/2 field, two massive spin-1 fields, and one massive spin-1/2
fermion.  This counting suggests that we must include at least
one chiral and one vector multiplet.  We will only consider
this minimal case in what follows.

The super-Higgs effect implies that after partial supersymmetry
breaking, the massive gravitino eats the
spin-1/2 Goldstone fermion.  A priori, we do not know whether the
Goldstone fermion is associated with the chiral or vector multiplet,
so we will let it be a linear combination of the two fields.

{}From this starting point, it is possible to construct the most general
supergravity coupling of the gravity and matter fields.  Recently,
we carried out this construction to lowest nontrivial order in
$1/v^2$ and $\kappa$, the inverse Planck mass \cite{BGO}.  We assumed
that supersymmetry is partially broken, that the cosmological constant
is zero, and that the second gravitino acquires a mass proportional
to $\kappa v^2$.

In our calculation, we demanded that the action be invariant under $N=2$
and central charge symmetry, and we required the transformations to
close (up to equations of motion on the fermions).  In the end,
we found the coupling to be unique.
As a check, we verified that in unitary gauge, our action
and transformations reduce to those of a massive $N=1$
spin-3/2 field, first worked out in ref.~\cite{FV}.

Our results show that the supergravity coupling of the Goldstone
system fixes the Goldstone fermion to be a particular linear
combination of the fermions from the chiral and vector multiplets.
The second supercurrent receives contributions from both fermions,
as well as from the second gravitino.

Once we derived the supercurrents, we used them
to compute the current algebra.  We found that the contribution
from the second gravitino exactly cancels the contributions from the
other two fermions, so the current algebra takes the following
form,
\be
\{\,Q_\alpha^A, \bJ_{m \dalpha B}\} = 2 \sigma^n_{\alpha\dalpha}
\,T_{mn} \,\delta^A{} _B.
\label{current alg 2}
\ee

In the presence of supergravity, there is no confusion
about the stress-energy tensor.  There is just one
such tensor, and it shows up on the right-hand side of the current
algebra.

How, then, is the no-go argument avoided?  Our results indicate that
the negative-norm components of the gravitino invalidate one of
the main assumptions behind the argument.  There is
no obstacle to partial supersymmetry breaking in the presence of
gravity.  The connection between this result and the membranes/D-branes
of rigid supersymmetry breaking is, at present, an urgent and
open question.

It is a pleasure to thank my collaborators, Sasha Galperin and Sam
Osofsky, for their insights on the work presented here.
I would also like to express my debt to Victor Ogievetsky for teaching
me the nonlinear approach to spontaneously broken spacetime symmetries.


\begin{thebibliography}{9}
\bibitem{Witten} E.  Witten,  Nucl.  Phys. B188 (1981) 513.
\bibitem{HLP}J.  Hughes, J.  Liu and J.  Polchinski,
Phys.  Lett. 180B (1986) 370;
J.  Hughes and J.  Polchinski, Nucl.  Phys.
B278 (1986) 147.
\bibitem{APT et al} I. Antoniadis, H. Partouche and T. Taylor,
Phys. Lett. B372 (1996) 83;
S. Ferrara, L. Girardello and M. Porrati, Phys. Lett. B376
(1996) 275.
\bibitem{others 1} M. Awada, M. Duff and C. Pope, Phys. Rev.
Lett. 50 (1983) 294; M. Duff, B. Nilsson and C. Pope,
Phys. Rep. 130 (1986) 1.
\bibitem{others 2}
S. Ferrara, L. Girardello and M. Porrati, Phys. Lett. B366
(1996) 155;
P. Fr\'e, L. Girardello, I. Pesando and M. Trigiante,
hep-th/9607032;
M. Porrati, hep-th/9609073, and references therein.
\bibitem{CWZ} S.  Coleman, J.  Wess and B.  Zumino,
Phys.  Rev.  177 (1969) 2239.
\bibitem{Volkov} D.V.  Volkov, Sov. J. Particles and Nuclei 4
(1973) 3;
V.I.  Ogievetsky, Proceedings of X-th Winter School of Theoretical
Physics in Karpacz, vol. 1, p. 227 (Wroclaw, 1974).
\bibitem{BG1} J.~Bagger and A.~Galperin, Phys. Lett. B336 (1994) 25.
\bibitem{BG2} J.~Bagger and A.~Galperin, hep-th/9608177, to appear
in Phys. Rev. D.
\bibitem{FV} S. Ferrara and P. van Nieuwenhuizen, Phys. Lett. B127 (1983) 70.
\bibitem{BGO} J.~Bagger, A.~Galperin and S.~Osofsky, to appear.
\end{thebibliography}
\end{document}